# Shear Deformation of Nonmodulated Ni₂MnGa Martensite: An *Ab Initio* Study

Martin Heczko[a*], Petr Šesták[b], Hanuš Seiner[c], Martin Zelený[a]

[a] Institute of Materials Science and Engineering, Faculty of Mechanical Engineering, Brno University of Technology, Technická 2896/2, 616 69 Brno, Czech Republic

[b] Institute of Physical Engineering, Faculty of Mechanical Engineering, Brno University of Technology, Technická 2896/2, 616 69 Brno, Czech Republic

[c] Institute of Thermomechanics, Czech Academy of Sciences, Dolejškova 5, 182 00 Prague, Czech Republic

∗ Corresponding author. E-mail address: martin.heczko1@ vutbr.cz

**Abstract**

The impact of shear deformation in $(101)[10\bar{1}]$ system of non-modulated (NM) martensite in Ni$_2$MnGa ferromagnetic shape memory alloy is investigated by means of *ab initio* atomistic simulations. The shear system is associated with twinning of NM lattice and intermatensitic transformation to modulated structures. The stability of the NM lattice increases with increasing content of Mn. The most realistic shear mechanism for twin reorientation can be approximated by the simple shear mechanism, although the lowest barriers were calculated for pure shear mechanism. The energy barrier between twin variants further reduces due to spontaneous appearance of lattice modulation or, in other words, the nanotwins with thickness of two atomic planes. Such nanotwins appear also on the generalized planar fault energy (GPFE) curve calculated using a newly developed advanced procedure and exhibits even lower energy than the defect free NM structure. These nanotwin doublelayers are also basic building blocks of modulated structures and play an important role in intermartensitic transformation.

1. **Introduction**

The Ni-Mn-Ga ferromagnetic shape memory alloys are under high scientific attention, as they are promising for various engineering applications [1, 2]. These alloys exhibit unique combination of exceptional mechanical properties [3–5], including very low twinning stress around 0.1 MPa [6], and ferromagnetic ordering with relatively high magnetic anisotropy [7]. Such combination of properties results in possible spontaneous macroscopic deformation of the alloy in an external magnetic field, which is known as magnetic-field-induced strain (MFIS) [8]. Depending on exact composition, these alloys are able to reach MFIS up to 12 % [6, 9]. The mechanism responsible for such a high strain is called magnetically induced reorientation (MIR) and is based on ability of martensitic twins to more likely change the orientation of their variant due to very low twinning stress rather than change the orientation of magnetic moments in crystal lattice due to high magnetic anisotropy [1, 10].

Practical utilization of Ni-Mn-Ga specific properties is significantly limited by operational temperature, which requires to be below the martensitic transformation temperature and the Curie temperature, while both these temperatures are very sensitive to small changes in chemical composition. Increasing content of Mn rises the martensitic transformation temperature on the one hand, but on the other hand, it decreases the Curie temperature in comparison to stoichiometric $Ni_2MnGa$ [11]. A large concertation of excess Mn even results in precipitation of a new phase. [12–14]. Exact structure of low-temperature martensitic phase is also highly dependent on composition [1, 10]. Far from stoichiometry [15, 16] the simplest martensitic structure occurs, the tetragonal non-modulated structure (NM), which exhibits too high twinning stress to enable MIR [17, 18]. The crystal lattice of NM martensite (see Fig. 1a) can be described as tetragonally distorted $L2_1$ cubic structure of parent phase, austenite, which is stable at elevate temperature. The MIR in NM martensite was reported only for alloys doped by additional elements [6, 9, 19]. Simultaneous addition of Co and Cu results in a decrease of $c/a$ tetragonal ratio and change of the orientation of the softest elastic shearing modes into the directions directly related to the MIR effect. Consequently, the mobility of $a/c$ twins in this alloy is increased and the MIR effect becomes enabled [20].

Other martensitic structures stable at and near stoichiometric compositions exhibit modulations of certain lattice planes. One of these structures is 10M martensite which exhibits nearly tetragonal structure with very small monoclinic angle and periodicity of modulation equal to five lattice planes. However, the modulation periodicity changes from commensurate to incommensurate with decreasing temperature which is accompanied by transition of monoclinic structure to tetragonal [21]. The crystal structure of 10M martensite results in complex twinned microstructure with twins at different length scales, starting from the nanoscale up to the macroscale [22, 23]. Such a hierarchical "twins-within-twins" microstructure seems to be directly related to very low twinning stress of a/c macrotwins and MFIS reported in 10M [22, 24]. The second modulated structure, the 14M martensite, has monoclinic lattice with monoclinic angle wider than reported for 10M and periodicity of modulation equal to seven lattice planes. Possible intermartensitic transformations were reported from 10M to 14M martensite and from 14M to NM martensite induced by changes in temperature or applied stress [16, 25].

The modulated structures can be also viewed as nanotwinned lattice of NM martensite [26]. In this concept, the modulation is described by an alternating sequence of nanotwins constituted from (101) lattice planes of NM structure. In particular, the structure of 10M martensite can be denoted within Zhdanov notation as $(3\bar{2})_2$, because it is described by a nanotwin with thickness of three planes oriented in one direction and a nanotwin with thickness of two planes oriented in the opposite direction. To obtain the complete structure of 10M martensite, this pattern has to be repeated once again due to periodicity in sublattices. Similarly, 14M martensite can be described as $(5\bar{2})_2$, i.e. the oppositely oriented nanotwins with thickness of five and two lattice planes also repeated twice (see [25, 27–29] for more details).

The last modulated structure we had to consider is 4O modulated martensite [30]. Using the nanotwinning concept, the 4O martensite can be described by the oppositely oriented nanotwins with thickens of two lattice planes, i.e. it could be denoted as $(2\bar{2})_1$. This structure is known to exist in the Ni-Mn-Sb(Co) alloys [31], and, although it has never been reported in experimental work for Ni-Mn-Ga, it occurs in *ab initio* calculations as a structure with the lowest energy for this material. However, similarly as for all other modulated structures, the stability of 4O further decreases with increasing content of Mn [32]. Several theories have been proposed to explain the lack of experimental observations of 4O martensite. According to the theory of adaptive martensite [26] the 4O does not fulfil adaptivity on habit plane with parent structure of austenite [28]. *Ab initio* calculations also found a barrierless, steeply descending transformation path only between austenite and 10M and 14M martensites, whereas transformations to other martensitic structures including 4O exhibit more or less significant barriers [29, 33]. It hinders the formation of 4O martensite from point of view of transformation kinetics. The lowest energy of 4O martensite provided by *ab initio* calculations can be also explained by overestimated electron delocalization in commonly used approximation for exchange-correlation potential [34, 35]. Employing the electron localization correction results in a higher stability of experimentally observed 10M structure than 4O structure [36].

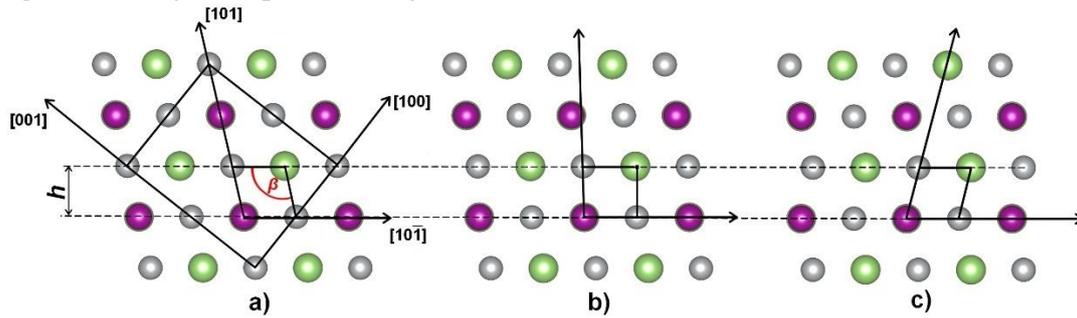

Fig. 1 (a) Tetragonally deformed L2$_1$ structure of NM martensite in $[100][001]$ coordinates can be described as monoclinic structure within $[101][10\bar{1}]$ coordinates, where $\beta$ is the shearing angle. (b) intermediate state between defect free structure and intrinsic stacking fault (c) fully developed intrinsic stacking fault after localized reorientation of the structure by simple shear mechanism. Ni, Mn and Ga atoms are marked grey, purple and green colors, respectively. The interplanar distance between (101) is denoted as *h*.

As follows from the nanotwinning concept, the reorientation of a/c twin variants in NM martensite and intermartensitic transformation between modulated and NM structures could be based on the same mechanism. The difference lies only in thickness of individual reoriented twins, which comprises from tens or hundreds of micrometers for the first, but only from a few of atomic layers for the second. Continuous development of twin on the atomic level and corresponding section of twinning energy landscape can be described by the generalized planar fault energy (GPFE) curve [37]. Within this model, twins are nucleated by intrinsic stacking fault (see Fig. 1) and propagate by gradual piling planar faults up one on another, creating peaks and valleys on energy profile reflecting the most stable positions of these defects and energy barriers, which must be overcome during movement of such planar faults. The movement of suitable planes itself is provided by local shearing of original lattice [38]. Therefore, the response of the lattice on shear deformation is crucial for understanding of the twinning mechanism. Probability of twin nucleation and propagation is dependent mainly on the height of energy barrier between individual steps along GPFE curve, but also on shear modulus and Poisson's ratio of twinned lattice [39]. Using this approach, Wang *et al.* predicted the lowest twinning stress for 10M Ni$_2$MnGa among other magnetic shape memory alloys. Surprisingly, this work does not include NM martensite.

The purpose of this paper is to study the effect of shear deformation applied along the crystallographic plane corresponding to the twinning plane in NM martensite. Here we present a detailed *ab initio* study of energy barriers for various shear models, and the corresponding changes of crystal lattice during the deformation. Increasing content of Mn in Ga sublattice is considered as well. We also provide calculations of energy barriers for transformation paths between NM and modulated martensites and GPFE curve for stoichiometric Ni$_2$MnGa, using a newly developed advanced procedure to find the optimized minima and maxima along the GPFE curve. The presented results and methodology provide a detailed description of shear deformation for Ni-Mn-Ga alloys and could shed light on process of twin reorientation and intermartensitic transformation.

2. **Computational methods and models**

*Ab initio* calculations based on spin-polarized density functional theory described in this work were done using Vienna *Ab Initio* Simulation Package (VASP) [40, 41]. The orbitals were expanded in terms of plane waves with kinetic energy lower than 600 eV. The interaction between electrons and ions was described by projected augmented waves (PAW) potential [42], whereas the electron exchange and correlation energy were calculated by the generalized gradient approximation (GGA) within the Pedrew-Burke-Ernzerhof (PBE) formalism [43]. A $\Gamma$-point centered mesh was used for the first Brillouin zone sampling with the smallest allowed spacing between **k**-points equal to 0.08 Å$^{-1}$. The first order Methfessel-Paxton smearing method with distribution parameter of 0.2 was used [44]. The geometry optimization was performed until atomic forces met their convergence criterion being lower than 1 meV · Å$^{-1}$. The effect of spin-orbit coupling is not considered in the described calculations. To reduce computational error caused by different cell sizes, the total energies are always plotted with respect to the undeformed cell with the same number of atoms as the deformed one. Convergence tests confirmed that the calculation settings are sufficient for estimating the total energy with a precision better than 0.2 meV/atom.

The reorientation of martensite via a/c twinning is an affine shear deformation of the lattice by applying shear strain in the $(101)[10\overline{1}]$ shear system. Therefore, we decided to use a monoclinic supercell within $[101][010][10\overline{1}]$ coordinates, substituting [100] for $[10\overline{1}]$ and [001] for [101] in the way depicted on Fig. 1. Monoclinic angle $\beta$ can be expressed as a function of $c/a$ tetragonal ratio as: $\beta = 2 \cdot arctan(c/a)$. The computational cell for simulation of shear deformation in stoichiometric Ni$_2$MnGa alloy contains eight atoms in two (101) planes, as shown on Fig. 1a. To model the off-stoichiometric compositions Ni$_{50}$Mn$_{28.125}$Ga$_{21.875}$ and Ni$_{50}$Mn$_{31.25}$Ga$_{18.75}$ the size of computational cells was increase to 2×1×2 and 1×1×2, respectively, and a single Ga atom was replaced by a Mn atom with opposite orientation of magnetic moments than remaining Mn and Ni atoms. During simulation, a shear deformation on (101) plane and $[10\overline{1}]$ direction was applied on whole computational cell to get the total energy profile and stresses along the deformation path. Here, initial NM martensite was deformed in 20 steps, until the monoclinic angle reached 90°. A structural optimisation with respect to constrained degrees of freedom was done for every single structure generated from the previous optimized steps by decreasing the $\beta$. We assume that the deformation path is symmetric around $\beta = 90°$ and should reach an oppositely oriented NM martensitic structure for $\beta < 90°$ as can be seen on Fig. 1 for case where only a part of the lattice is reoriented locally. For complete description of shear modes, specific degrees of freedom needed to be constrained. For this purpose, the advanced optimization with help of delocalized internal coordinates implemented in Gadget tool was employed [45].

Tab. 1 Summary of deformation models and related constrains of the computational cell.

| deformation mechanism | constrained cell parameters along the whole path | additional constrain in each single step on the path |
|---|---|---|
| pure shear | relative atomic coordinates, angles $\alpha$ and $\gamma$ | angle $\beta$ |
| twinning shear | cell size in shear plane, relative atomic coordinates, angles $\alpha$ and $\gamma$ | angle $\beta$ |
| simple shear | cell size in shear plane, height of the cell perpendicular to shear plane, relative atomic coordinates, angles $\alpha$ and $\gamma$ | angle $\beta$ |
| pure modulated | angles $\alpha$ and $\gamma$ | angle $\beta$ |
| twinning modulated | cell size in shear plane, angles $\alpha$ and $\gamma$ | angle $\beta$ |
| simple modulated | cell size in shear plane, height of the cell perpendicular to shear plane, angles $\alpha$ and $\gamma$ | angle $\beta$ |
| non-optimized GPFE | cell size in shear plane, height of the cell perpendicular to shear plane, relative atomic coordinates, angles $\alpha$ and $\gamma$ | angle $\beta$ |
| optimized GPFE | cell size in shear plane, height of the cell perpendicular to shear plane, angles $\alpha$ and $\gamma$, atomic coordinates constrained by SS-NEB | none |
| NM→modulated transformation path | atomic coordinates and cell size constrained by SS-NEB | none |

By constraining different numbers of degrees of freedom (that is, fixing different lattice parameters), energy functions were obtained for three different shear mechanisms:

1. The smallest number of degrees of freedom are constrained in the pure shear mechanism (Fig. 2a). This mechanism represents the reorientation of martensite from one variant into another through evolution of the shear deformation angle β, without considering any additional constraints the surrounding material may impose onto the lattice. For each value of β, all other crystallographic parameters are assumed as variable and are obtained from structural relaxation in each step. However, the cell angles α and γ do not change from 90° within relaxation to keep the symmetry of the lattice. It is worth noting that the relaxation of lattice dimensions allows the unit cell to expand or shrink in the (101) twinning plane, violating, thus, the kinematic compatibility conditions on this plane. This means that while the pure shear mechanism might be the most energetically efficient (because of relaxing the highest number of parameters), it is geometrically not possible inside of a bulk crystal. In this sense, the energy function for this mechanism can be expected to be the lower bound for the real energy barrier.

2. In the second considered deformation model, the twinning shear (Fig. 2b), we disabled optimization of lattice angles, and the shearing plane defined by two perpendicular lattice vectors, constraining the cells length in [101] and [010] directions, i.e. the *a* and *c* lattice constants of the monoclinic cell. While vector fixation prevents changes in (101) twinning plane, interplanar distance between these planes can be relaxed as a reaction on the applied load. The twinning mode represents strains satisfying the kinematic compatibility conditions for a planar interface running along the (101) plane, that is, it can be linked to a jump in the macroscopic deformation gradient carried by such an interface moving through the crystal (see e.g. Ref. [46]). For this reason, the constraints on twinning plane and variability of other crystallographic parameters makes this mode the most suitable to describe the shear mechanism for twinning.

3. The last shear model, the simple shear (Fig. 2c), has lattice parameters *a* and *c* and crystallographic angles constrained, leaving only the atoms being able to optimize their positions during structural relaxation. Only *b* lattice vector shortens along simple shear deformation path, but its component perpendicular to shearing plane (height of the cell $h_b$) remains constant, which makes the interplanar distance unchanged. This mechanism obeys the geometrical constraints resulting from the compatibility conditions, and prevents, in addition, the lattice from shrinking in the direction perpendicular to the (101) twinning plane. Such conditions might apply for example for at faces of a thin needle of one variant of martensite growing into another. The additional constraint increases the energy along the reorientation path, and thus, the resulting energy function can be considered possibly as an upper bound for the real energy. Equivalent shear mechanisms can be assumed also for the modulated lattices, as outlined in Fig. 2d,e.

Then, simple shear mechanism was picked to simulate shearing during GPFE curve calculations. The monoclinic supercell for GPFE calculations was described by $[10\bar{1}][010][404]$ vectors of tetragonal lattice containing eight (101) planes. In this cell, the lower four atomic planes were fixed, while a displacement of $u = \frac{|\vec{s}|}{n_{GPFE}}$ was cumulatively applied to the upper four planes creating planar faults (see Fig. 1). The vector $\vec{s}$ is shearing vector in $[10\bar{1}]$ direction and $n_{GPFE}$ is number of deformation steps between expected GPFE minima corresponding stable twin configurations.

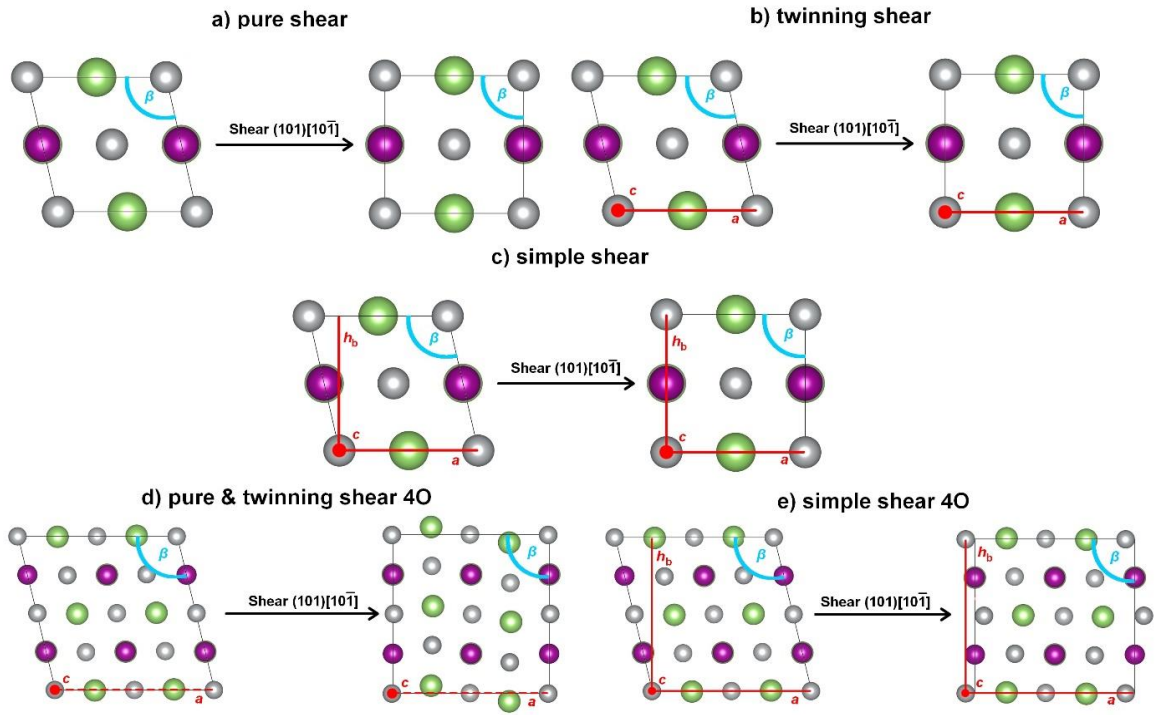

Fig. 2: Shear mechanisms considered in this paper: (a) pure shear with allowed relaxation in the a- and c-axis directions, (b) twinning shear respecting the geometry constraints coming from the compatibility conditions at the twinning plane, (c) simple shear mechanisms. (d) corresponds to pure and twinning shear mechanisms and (e) to simple shear mechanism in 2×1×2 supercell when formation of lattice modulation is allowed. Cell parameters constrained during entire deformation path are red, the additional parameters constrained in each individual deformation step but developed along entire path are blue. In (d) the red dashed line corresponds to constraints only for the twinning mechanism. The computational cell is viewed from [010] direction.

The shearing vector $\vec{s}$ then describes displacement of (101) planes necessary to form a perfectly mirrored twin in tetragonal lattice [47]. Once the local energy minimum was obtained, the movement of the plane closest to fixed part of crystal was stopped and the movement of the rest continued. Between each neighbouring minima a set of several total energy points along the deformation path was calculated. Lattice vectors and atomic positions were not optimized during this path.

After finding all local minima on the non-optimized GPFE curve, which is a result of above-mentioned shearing, the lattice angle $\beta$ and atomic positions were further optimized by Gadget tool [45] with constraints respecting the simple shear mechanism. To find optimized maxima, however, a different approach has been considered. Because the common approach for optimization of energy maxima leads to unrealistic structural changes and discontinuities on energy profile, the solid-state nudged elastic band (SS-NEB) method with constrained cell [48] implemented into Atomic Simulation Environment (ASE) [49] was used instead. This method serves to find transition states for transformation between initial and final structures based on several images linked to each other. The optimization is constrained by adding spring forces between images, preventing them from finding local minima, and is proceeded by projecting out the force component of due to the potential perpendicular to the band. Using the optimized neighbouring local minima as starting and final points with at least 3 images respecting geometrical constrains (Tab. 1) from simple shear model between them, our optimized GPFE curve was obtained. The SS-NEB approach but without any constrain was use also for estimation of energy barriers along NM→4O, NM→10M and NM→14M transformation paths. The final modulated structures of 4O, 10M and 14M martensites were created first with help of nanotwinning concept and then fully relaxed. The computational cells describe the modulated structures in so called diagonal coordinates (see Appendix in [25] for more details or [28, 29]), which allows to directly connect them with monoclinic description of NM lattice by shear deformation applied along basal plane, i.e., in the same shear system $(101)[10\bar{1}]$ of NM lattice as in previously described calculations. Then the monoclinic lattice angle $\beta$ can be use as transformation coordinate.

3. **Results**

In the first step we investigated the response of the whole crystal lattice to affine shear deformation in the $(101)[10\bar{1}]$ shear system. For stoichiometric alloy, a symmetric deformation path around 90° is expected, with formation of oppositely oriented sheared structure of NM martensite for $\beta = 180° - \beta_{NM}$ [50]. Total energy profiles as a function of angle $\beta$ along shear deformation paths are shown on Fig. 3(a) for this composition. As expected, the mechanisms with less degrees of freedom and more constrained structure optimization exhibit higher energetic barrier for reorientation. The total energy profile for pure shear mechanism (the black curve) is equivalent to the energy profile along tetragonal deformation path. It means that for $\beta = 90°$ the cubic structure of austenite is stabilised in local minimum. Reaching the state of austenite is restricted with conserving any dimension of the cell in simple shear and twinning shear mechanisms during deformation. The mechanism with the most constrained structural optimisation, the simple shear mechanism (the red curve), exhibits its global maximum of total energy $\Delta \mathrm{E}_{tot}^{simple}$ for $\beta = 90°$ approximately 15 meV/atom above NM martensite, which agrees with previously published data [50]. The energy maximum for twinning shear mechanism $\Delta \mathrm{E}_{tot}^{twinning}$ is only about 0.5 meV/atom lower (the blue curve) then for simple shear mechanism.

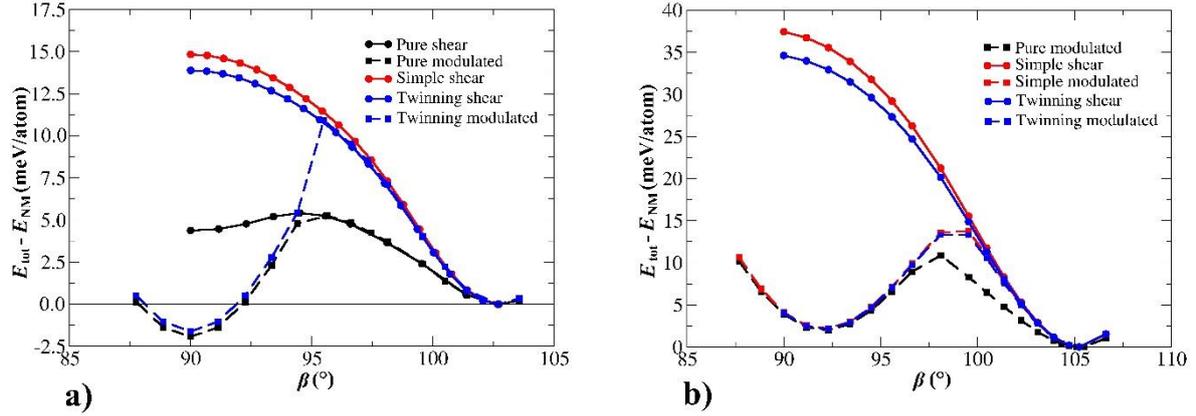

Fig. 3 Total energy profiles along different shear deformation paths (a) for stoichiometric Ni$_2$MnGa alloy, (b) for off-stoichiometric Ni$_{50}$Mn$_{31.25}$Ga$_{18.75}$ alloy.

As can be seen on Fig. 3(b), the energy barriers reach higher values for all shear mechanisms in the alloy with increased content of Mn from 25 at. % to 31.25 at. %. However, the off-stoichiometric alloy exhibits a different profile of total energy for the pure shear deformation with a very deep minimum near $\beta = 90°$, if relaxation of all atomic positions is allowed (black dashed curve, the mechanisms denoted as pure modulated). The minimum corresponds to a modulated structure with periodicity of modulation equal to four layers, i.e. the structure nearly identical with 4O martensite. The modulated structure is stabilized at $\beta \cong 92.3°$ for Ni$_{50}$Mn$_{31.25}$Ga$_{18.75}$ alloy. Moreover, the same minimum appears also on curves for other mechanism, if relaxation of all atomic positions is allowed (red and blue curve, the mechanisms denoted as simple and twinning modulated). Similar total energy profiles were obtained also for Ni$_{50}$Mn$_{28.125}$Ga$_{21.875}$ alloy (not shown). Only differences were that the minima corresponding to modulated structures could be found for $\beta \cong 91.0°$, and energy maxima along the paths were lower due to the lower content of excess-Mn. The corresponding values of $\Delta E_{tot}$ for all compositions can be found in Table 1 together with the highest stresses $\sigma_{max}$ found along the paths. The value of $\sigma_{max}$ corresponds to the inflection point on the total energy profile and can be understood as the ideal shear strength of NM lattice for deformation in (101)[10$\bar{1}$] shear system. Because the angle $\beta$ for energy minima corresponding to modulated structure differs from 90°, the modulated deformation paths for off-stoichiometric compositions are not perfectly symmetrical, due to decreased lattice symmetry caused by Mn-excess atoms in Ga sublattice. To avoid formation of modulated structure in twinning and simple shear mechanisms, the optimization of atomic position was disabled (blue and red curves, the mechanisms denoted as twinning and simple shear). Then the total energy profiles have similar shape to the profiles of corresponding twinning and simple shear mechanisms for stoichiometric composition on Fig. 3(a) but with significantly higher energies at maxima.

Although the formation of modulated structure on the shear deformation path is independent on the used shear mechanism, its orientation is determined by remaining degrees of freedom and size of the computational cell. When the (101) interplanar distance is not constrained in terms of shear deformation, and the cell contains four atomic planes in both [101] and [10$\bar{1}$] direction (the 2×1×2 cell originally used for Ni$_{50}$Mn$_{28.125}$Ga$_{21.875}$ composition), resulting 4O structure on the path prefers the nanotwinning planes oriented perpendicular to shear direction, i.e., the modulation vector is parallel with shear direction (see Fig 2(d)). On the other hand, the orientation of nanotwinning planes is parallel with shear direction for the mechanism denoted as simple modulated. Therefore, the nanotwinning planes and the shear plane are the same crystallographic planes, and the modulation vector is perpendicular to shear direction as can be intuitively expected (see Fig 2(e)). However, the orientation of resulting modulated structure do not have impact on the profile of total energy.

Due to exceptional stability of modulated structure in non-stoichiometric alloys, we considered to implement it into pure shear and twinning shear mechanisms for stoichiometric $Ni_2MnGa$. In stoichiometric composition the formation of modulated structure is disabled by smaller computational cell and higher symmetry of the lattice, which does not contain Mn atoms in Ga sublattice. The total energy profiles for mechanisms denoted as pure and twinning modulated in Fig. 3(a) (black and blue dashed curves) were obtained for increased computational cell (2×1×2 as for $Ni_{50}Mn_{28.125}Ga_{21.875}$) containing four atomic planes where a small symmetry breaking was introduced in each step before relaxation. Total energies along such paths are identical with energies along the shear paths, until the modulation starts developing at specific shear angle. Then, significant decrease of energy was observed leading to the energy minimum at $\beta = 90°$ which is about 2 meV/atom lower than NM structure. The same energy difference was reported between 4O and NM martensite in our previous work [30]. The $Ni_{50}Mn_{28.125}Ga_{21.875}$ alloy exhibits still lower energy of modulated structure on the shear path than NM structure, but just about -0.5 meV/atom (not shown). The modulated structure becomes less stable than NM for $Ni_{50}Mn_{31.25}Ga_{18.75}$ alloy as can be seen on Fig. 3(b). It confirms decreasing stability of modulation with increasing content of Mn as reported in previous calculations and experiments [7, 15, 16, 30, 32].

Energy profiles for all described shear mechanisms show a strong tendency of the lattice to form modulated structures if relaxation of all atomic positions is allowed. However, these models are limited by size of the used computational cell, which does not allow full development of the structure due to periodic boundary conditions to reach the lowest energy. Because the computational cell has only four atomic planes, the modulation periodicity of resulting structure is restricted to form the 4O structure only. For this reason, we employed SS-NEB method to calculate the minimum energy path for transformations between NM and modulated structures in stoichiometric composition and determine the most reliable path for intermartensitic transformation with the lowest energy barrier. The minimum energy paths as a function of monoclinic angle $\beta$ are shown on Fig. 4.

The NM→4O transformation shows the same profile of total energy as the profile obtained by the pure modulated mechanism in Fig. 3(a), because the SS-NEB calculations of transformation paths do not consider any lattice constraint. Moreover, the energy profile along the path seems to be not dependent on the orientation of final 4O structure, because in all SS-NEB calculation presented in this work, the nanotwinning planes in resulting structures are oriented parallel with the shear system $(101)[10\bar{1}]$. Transformation path between NM and 10M martensite exhibits a similar barrier as for NM→4O. The barrier lower about ~1 meV/atom was found for NM→14M transformation path. It suggests that NM martensite will preferably transform to 14M than for 10M or even to 4O. This finding really agrees with experimental observation [25], because the commonly observed sequence of intermartensitic transformation is 10M↔14M↔NM [51–53]. To further reveal the ability of NM lattice to form twins and nanotwins, we also calculated the GPFE curve for stoichiometric $Ni_2MnGa$ employing the constraints corresponding to the simple shear model, i.e. the cell size in [101] and [010] directions and interplanar distance between (101) remain unchanged.

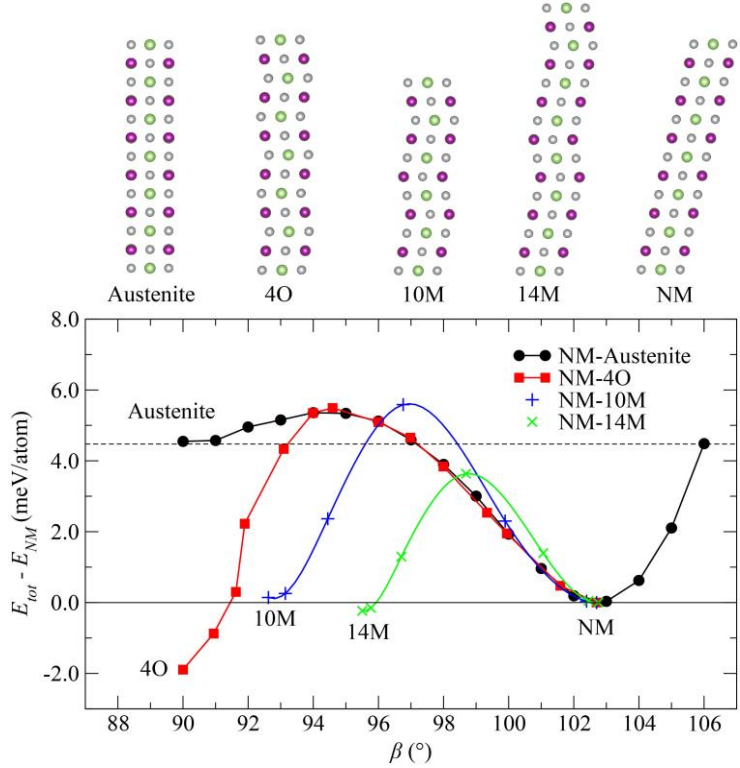

Fig. 4 Total energies along transformation paths between NM martensite and modulated structures for stoichiometric Ni$_2$MnGa calculated by SS-NEB.

The solid line on Fig. 5 corresponds to GPFE obtained without structural optimization. The curve consists of several maxima and minima. The first minimum corresponds to the energy of intrinsic stacking fault, i.e. one-layer twin, whereas other minima correspond to twins with increasing thickness, i.e. twin with thickness of two atomic layers, with thickness of three atomic layers etc. The first maximum is then a barrier which has to be overcome to create intrinsic stacking fault and other maxima are barriers for further growth of the twin. Usually only the first maximum and intrinsic stacking fault exhibit different energies than other maxima and minima on a GPFE curve, which means that once a stacking fault is created the energy of twin boundaries and the further growth of the twin is independent of its thickness. In comparison with GPFE curves published for other materials [47, 54] including also shape memory alloys [37, 55], Ni$_2$MnGa does not exhibit such behaviour. Here, the energy of twin is highly dependent on twin thickness as well as on the barriers for growing of a twin. The even minima are generally lower than the odd ones.

Moreover, the energy barrier between intrinsic stacking fault and two-layer twin is significantly smaller than other barriers on the GPFE curve. The effect is even more pronounced if structural optimization is allowed. The energy of two-layer twin decrease to -0.24 meV/atom and the energy of fourth layered twin to -0.16 meV/atom, which indicates their stabilization in comparison to the defect free structure. In other words, the energy of nanotwin boundaries is negative for such arrangement of atomic planes. The least obvious effect of structural optimization is seen in the first barrier, which energy decreases from 4.42 meV/atom to 3.90 meV/atom. For other maxima and minima along the curve, the decrease in energy is more or less constant. Therefore, the relative height of barriers is not affected significantly by structural optimization. Comparison to the data revealed in [33] shows that our values for minima along the GPFE curve are generally lower for around 0.5 meV/atom than previously published values.

Since the energy of symmetrical twin is dependent on total width of such twinned layer [33], we also compared the energy of $(4\bar{4})_1$ nanotwin to twice wider $(8\bar{8})_1$ nanotwin to be sure about sufficiency of supercell containing only 8 layers, both in optimized state. Then the interface energy of $(8\bar{8})_1$ nanotwin corresponds to GPFE energy of $(4\bar{4})_1$ nanotwin with accuracy around 0.05 meV/atom which confirms a reliability of the choose $(4\bar{4})_1$ twinned model.

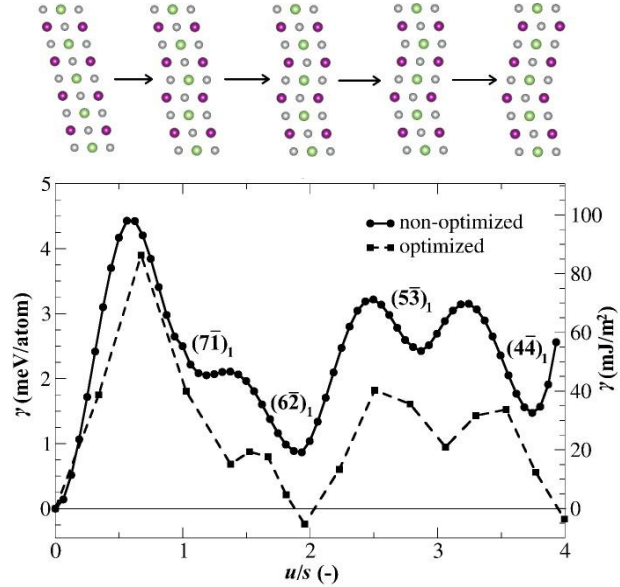

Fig. 5 GPFE curve for stoichiometric Ni$_2$MnGa, i.e. the total energy per atom with respect to energy of defect free NM martensite as a function of shear displacement relative to the magnitude of shear vector $\vec{s}$ [45]. Alternative axis displays the corresponding energy of planar fault. Nonoptimized curve is demonstrated by solid line, while the optimized is depicted with dashed line.

## 4. Discussion

Since there are not any constrained dimensions in pure shear model, the shear deformation path is comparable to the tetragonal compression of the lattice along the [001] direction. The height of energetic barriers between austenite and martensite for these two deformations are also identical. Although the pure shear gives the lowest values of the energy barrier and the lowest shear stress for reorientation of the structure from one twin variant of NM martensite to another, the probability of its existence in a real material is rather low. The reason lays in dimensional changes along the twinning plane needed for its successful propagation, which contradicts the compatibility constrains on twinning plane. As the volume corresponding to reoriented lattice is relatively low, surrounding material would have to react on any changes of volume or lattice parameters, which may increase energy of the whole system.

This fact makes twinning shear model the most realistic one, as the twinning plane conservation is included. Such model allows the material to reorientate via a motion of a single planar twin interface. In experimentally observed twinning, the twinning plane does not undergo any changes during shear deformation leading to twin reorientation. [56, 57] Furthermore, it was found out, that twinning and simple shear models are similar in the terms of total energy, although the values of shear stress are slightly higher and maximal stresses slightly lower for simple shear mechanism (see Table 2). However, there are still some remaining degrees of freedom in the twinning shear model, among which the simple shear model is the easiest to apply and also more time-efficient [58, 59]. Moreover, the recent experimental study [25] reveals that interplanar distances between (101) plane in NM martensite and (110) plane in modulated structure (i.e. the nanotwinning plane, described in so called cubic coordinates) remain unchanged within the intermartensitic transformation, i.e. within the reorientation of nanotwins, if nanotwinnig description of the modulated lattice is adopted. Therefore, this mechanism seems to be sufficient for description of twin reorientation and calculations of GPFE curves.

Tab. 2 Comparison of total energy maxima (Δ$E_{tot}$) and shear stress maxima ($σ_{max}$) along the shear deformation paths. Values in parentheses were obtained for paths where formation of modulated structure was not allowed.

| Alloy | $\Delta E_{tot}^{pure}$ [meV/atom] | $\Delta E_{tot}^{twinning}$ [meV/atom] | $\Delta E_{tot}^{simple}$ [meV/atom] |
|---|---|---|---|
| Ni$_2$MnGa | 5.25 (5.42) | 9.59 (13.88) | (14.83) |
| Ni$_{50}$Mn$_{28.125}$Ga$_{21.875}$ | 6.63 | 10.28 (23.11) | 10.72 (24.52) |
| Ni$_{50}$Mn$_{31.25}$Ga$_{18.75}$ | 10.88 | 13.32 (34.63) | 13.77 (37.45) |
| Alloy | $\sigma_{max}^{pure}$ [GPa] | $\sigma_{max}^{twinning}$ [GPa] | $\sigma_{max}^{simple}$ [GPa] |
| Ni$_2$MnGa | 0.78 (0.68) | 1.53 (1.54) | (1.59) |
| Ni$_{50}$Mn$_{28.125}$Ga$_{21.875}$ | 1.08 | 1.93 (2.15) | 2.00 (2.20) |
| Ni$_{50}$Mn$_{31.25}$Ga$_{18.75}$ | 1.86 | 2.41 (2.82) | 2.48 (3.00) |

Our calculations show excessive stability of modulation within the shear deformation of NM martensite even in non-stoichiometric compositions. Appearance of the 4O structure ($(2\bar{2})_1$ in nanotwinning notation) along the shear deformation paths is, nevertheless, enforced by the size of computational cell. If the computational cell is too small, the modulation does not appear, as can be seen on Fig. 3(a) where the cell contains only two layers for pure, twinning and shear paths. On the other hand, using a large cell will results in a different type of modulation described by appropriate sequence nanotwins, however, always with preference for nanotwins with thickness of two planes as can be seen on Fig. 5. Importance of this nanotwin doublelayer for stability of modulated structure has been discussed in our early work [30] and explained by Gruner *et al.* by frustrated antiferromagnetic exchange in NM structure, which increases its energy compared to 4O [33]. Current work further confirms this idea as the doublelayers form the modulated structures along the shear path and also the twin with thickness of two atomic planes exhibits the lowest energy on GPFE curve. This structure containing nanotwin doublelayer can be also denoted as $(6\bar{2})_1$ using the notation arising from nanotwinning concept, because the computational cell used for GPFE calculations consist of eight atomic layers. Thus, the whole sequence of structures corresponding to minima along the curve is: NM→ $(7\bar{1})_1$ → $(6\bar{2})_1$ → $(5\bar{3})_1$ → $(4\bar{4})_1$, which explains the increased stability of twins with thickness of odd number of layers. Different stability of twins with even and odd number of atomic planes and negative energy of nanotwin boundaries with specific distance make the application of the model for prediction of twinning stress proposed by J. Wang and H. Sehitoglu [37] difficult in case of Ni$_2$MnGa NM martensite. Comparison of energy barrier along the transformation paths between NM and 14M structures and the first energy barriers on the GPFE curve (the barrier for creation of intrinsic stacking fault) reveals they are of nearly the same height corresponding to approximately 4 meV/atom. It suggests that shear induced intermartensitic transformation can be realized by subsequent steps comprising of local reorientation of nanotwin doublelayers. The higher barriers reported for NM→10M and NM→4O indicate that a sufficiently large number of atomic layers between doublelayers decreases the barrier. In 4O $((2\bar{2})_1)$ and 10M $((3\bar{2})_2)$, there are two and three atomic planes, respectively, between doublelayers, whereas in 14M $((5\bar{2})_2)$ and in low-energy structure $(6\bar{2})_1$ on GPFE curve, the doublelayers are separated by five and six atomic planes.

The excessive stability of modulated structures and nanotwin doublelayer along the shear deformation paths also suggests that the reorientation of one variant of macroscopic twin to another variant could be mediated by local creation of nanotwins (local modulation) in the vicinity of twinning plane. However, such behaviour was not observed experimentally, probably because the nanotwins could exist only for very short time in an intermediate state between both twin variants. Missing experimental evidence could be also explained by fact, that NM is not stable for stoichiometric composition and twin reorientation in modulated structures is based on more complex mechanism than simple shearing of planes [60, 61] . The reorientation of twins in offstoichiometric alloys with stable NM martensite is rarely studied due to the high twinning stress [7, 62]. On the other hand, application of external load on such alloys will rather results in the formation of modulated structure within the shear induced intermartensitic transformation than in reorientation of macroscopic twins [63, 64]. Recently, M. Vronka *et al.* [65] observed a modulation in thin lamellae for HRTEM even in alloy doped by Co and Cu, which otherwise exhibits high stability of NM structure and for which low twinning stress was reported [9]. The observed modulation is explained by its association with the stress in the foil which indicates that studied alloy is on the edge of the shear instability of NM structure [65]. This agrees with our calculations showing that, modulation spontaneously appears in NM martensite when it is exposed to applied shear load.

The increasing Mn content in the alloy significantly increases the stability of NM structure, which has been confirmed by many previous theoretical and experimental investigation [15, 16]. The stabilizing effect of Mn is noticeable from the increasing values of energy barrier along the shear paths $\Delta E_{tot}$, and from the ideal shear strength $\sigma_{max}$, which represents the upper limit of the shear stress that can be applied to the material without irreversible damage. In reality, other deformation mechanisms take place than simultaneous shearing of all bonds along a given crystallographic plane, which significantly reduce the necessary stress. This effect can be seen in decreasing of $\sigma_{max}$ when creation of modulation is allowed by size of the cell. As the value of $\sigma_{max}$ corresponds to the inflection point on the total energy profile, it allows to estimate also the maximal shear strain $\beta_{max}$ which can be applied before breaking of interplanar bonds or for which the elasticity of a material behaves linearly. In stoichiometric alloy $\beta_{max}$ corresponds approximately to 5.5 % of engineering strain independently on shear mechanism with exception for pure modulated path. Here $\beta_{max}$ is equal to 3.8 %. These values are significantly lower than the values reported for pure metals and other materials, where $\beta_{max}$ reach the values between 10 % and 35 % [66–69]. On the other hand, the values of $\sigma_{max}$ are similar to values reported for soft and ductile metals like gold, silver or copper. With an increasing content of Mn, $\beta_{max}$ also increases to values of approximately 7.6 % for $Ni_{50}Mn_{28.125}Ga_{21.875}$ alloy and 9.7 % for $Ni_{50}Mn_{31.25}Ga_{18.75}$ alloy, which indicates that alloys with high content of Mn behave more like common materials. Even lower value of $\beta_{max}$ then for pure modulated path in $Ni_2MnGa$ was found for NM→14M transformation. Here $\beta_{max}$ is equal to 2.9 %. Therefore, a very small load has to be introduced to Ni$_2$MnGa NM martensite to initiate formation of nanotwins.

**Conclusions**

Using *ab initio* atomistic simulations based on DFT together with three different deformation models (lattice constrains), we investigate the impact of shear deformation on NM martensite of Ni$_2$MnGa magnetic shape memory alloy. In particular, the shear system $(101)[10\bar{1}]$ was chosen for investigation because it corresponds to NM lattice twinning and can be also associated with shear induced intermartensitic transformation from NM to modulated martensite, adopting the nanotwinning description of modulated lattices. We found the twinning shear mechanism the most realistic one for description of reorientation of twin variants, although the pure shear mechanism exhibits lower energy barrier along the shear path. However, the twinning shear mechanism can be successfully approximated by the simple shear mechanism, which is easy to implement and increase the efficiency of calculation with comparable accuracy. It was also demonstrated that the choice of the deformation model and the size of the simulation space play an essential role.

The stability of the NM lattice with respect to shear increases with increasing content of Mn in the alloy. Compared to pure metals and other common materials, the Ni$_2$MnGa NM martensite exhibits significantly lower value of maximal shear strain $\beta_{max}$, the strain which can be applied to a material before irreversible changes appears.

We also found that the lattice modulations or, in other words, the formation of nanotwins appears spontaneously along the deformation path. It significantly reduces the energy barrier, which has to be overcome to reach the target orientation of the twin variant. Although the periodicity of modulation is enforced by the size of computational cell, we observe a common feature which is the formation of nanotwins with thickness of two atomic planes. Such arrangements of nanotwins also appear on the GPFE curve as the minimum with the lowest energy, even lower than for the defect free NM structure. Such irregularity on GPFE curve makes it difficult to apply the established theories for twin motion and prediction of twinning stress. These nanotwin doublayers are also basic building blocks of modulated structures [70] and play an important role in intermartensitic transformation. The barrier for its formation is lowered in case where the doublelayers are separated by five and more atomic planes. It makes the direct transformation from NM to 14M structure favourable compared to transformations to 10M or 4O structures.

**Acknowledgements**

This work was financially supported by the Czech Science Foundation grant no. 21-06613S. Computational resources were provided by the Ministry of Education, Youth and Sports of the Czech Republic under the Projects e-INFRA CZ (ID:90140) at the IT4Innovations National Supercomputing Center. Figures 1, 2, 4 and 5 were visualized using the VESTA software [71] (version 3, National Museum of Nature and Science, 4-1-1, Amakubo, Tsukuba-shi, Ibaraki 305-0005, Japan). H. Seiner acknowledges the support from the Operational Programme Johannes Amos Comenius of the Ministry of Education, Youth and Sport of the Czech Republic, within the frame of project Ferroic Multifunctionalities (FerrMion) [project No. CZ.02.01.01/00/22_008/0004591], co-funded by the European Union.